\documentstyle[12pt]{article}

\begin{document}

\baselineskip=16pt plus 1pt minus 1pt

\begin{center}{\large \bf 
Z(5): Critical point symmetry for the prolate to oblate 
nuclear shape phase transition
} 

\bigskip\bigskip

{Dennis Bonatsos$^{\#}$\footnote{e-mail: bonat@inp.demokritos.gr},
D. Lenis$^{\#}$\footnote{e-mail: lenis@inp.demokritos.gr}, 
D. Petrellis$^{\#}$, 
P. A. Terziev$^\dagger$\footnote{e-mail: terziev@inrne.bas.bg} }
\bigskip

{$^{\#}$ Institute of Nuclear Physics, N.C.S.R.
``Demokritos''}

{GR-15310 Aghia Paraskevi, Attiki, Greece}

{$^\dagger$ Institute for Nuclear Research and Nuclear Energy, Bulgarian
Academy of Sciences }

{72 Tzarigrad Road, BG-1784 Sofia, Bulgaria}

\end{center}

\bigskip\bigskip
\centerline{\bf Abstract} \medskip
A critical point symmetry for the prolate to oblate shape phase transition 
is introduced, starting from the Bohr Hamiltonian and approximately separating
variables for $\gamma=30^{\rm o}$. Parameter-free (up to overall scale factors)
predictions for spectra and B(E2) transition rates are found to be in good 
agreement with experimental data for $^{194}$Pt, which is supposed to be 
located very close to the prolate to oblate critical point, as well as 
for its neighbours ($^{192}$Pt, $^{196}$Pt). 

\bigskip\bigskip
PACS numbers: 21.60.Ev, 21.60.Fw, 21.10.Re 

Keywords: Z(5) model; Critical point symmetry; Shape phase transition; 
Prolate to oblate transition; Triaxial rotator  

\newpage

{\bf 1. Introduction}

Critical point symmetries in nuclear structure are recently receiving
considerable attention \cite{IacE5,IacX5,IacY5}, 
since they provide parameter-free (up to overall
scale factors) predictions supported by experimental evidence
\cite{CZE5,Zamfir,CZX5,Kruecken}.
So far the E(5) [U(5) (vibrational) to O(6) ($\gamma$-unstable)] 
\cite{IacE5,CZE5,Zamfir} and the X(5) [U(5) to SU(3) (prolate deformed)] 
\cite{IacX5,CZX5,Kruecken} critical point symmetries have been considered, 
with the recent addition of Y(5) \cite{IacY5}, related to the transition 
from axial to triaxial shapes. All these critical point symmetries have 
been constructed by considering the original Bohr equation \cite{Bohr}, 
separating the collective $\beta$ and $\gamma$ variables, and making different 
assumpions about the $u(\beta)$ and $u(\gamma)$ potentials involved.    
 
Furthermore, it has been demonstrated \cite{Jolie68} that experimental data 
in the Hf-Hg mass region indicate the presence of a prolate to oblate shape 
phase transition, the nucleus $^{194}$Pt being the closest one 
to the critical point. No critical point symmetry for the prolate to oblate 
shape phase transition originating from the Bohr equation has been given 
so far, although it has been suggested \cite{Jolie87,Jolie89}
that the (parameter-dependent)
O(6) limit of the Interacting Boson Model (IBM) \cite{IA} can serve as 
the critical point of this transition, since various physical quantities 
exhibit a drastic change of behaviour at O(6), as they should \cite{Werner}.  

In the present work a parameter-free (up to overall scale factors) critical 
point symmetry, to be called Z(5), is introduced for the prolate to oblate 
shape phase transition, leading to parameter-free predictions which compare 
very well with the experimental data for $^{194}$Pt. The path followed 
for constructing the Z(5) critical point symmetry is described here:

1) Separation of variables in the Bohr equation \cite{Bohr} is achieved by 
assuming $\gamma=30^{\rm o}$. When considering the transition from 
$\gamma=0^{\rm o}$ (prolate) to $\gamma=60^{\rm o}$ (oblate), it is 
reasonable to expect that the triaxial region 
($0^{\rm o} < \gamma < 60^{\rm o}$) will be crossed, 
$\gamma=30^{\rm o}$ lying in its middle. Indeed, there is experimental 
evidence supporting this assumption \cite{Gizon}. 

2) For $\gamma=30^{\rm o}$ the $K$ quantum number (angular momentum projection 
on the body-fixed $\hat z'$-axis) is {\sl not} a good quantum number any more,
but $\alpha$, the angular momentum projection on the body-fixed $\hat x'$-axis 
is, as found \cite{MtVNPA} in the study of the triaxial rotator 
\cite{DavFil,DavRos}. 

3) Assuming an infinite well potential in the $\beta$-variable and a 
harmonic oscillator potential having a minimum at $\gamma=30^{\rm o}$ 
in the $\gamma$-variable, the Z(5) model is obtained. 

On these choices, the following comments apply: 

1) Taking $\gamma=30^{\rm o}$ does not mean that rigid triaxial shapes are 
prefered. In fact, it has been pointed out \cite{ZC260} that a nucleus in a 
$\gamma$-flat potential \cite{Wilets}
(as it should be expected for a prolate to oblate 
shape phase transition)  oscillates uniformly over $\gamma$ from 
$\gamma=0^{\rm o}$ to $\gamma=60^{\rm o}$, having an average value 
of $\gamma_{av}=30^{\rm o}$, and, therefore, the triaxial case to which 
it should be compared is the one with $\gamma=30^{\rm o}$. 
Furthermore, it is known \cite{Casten} that many predictions of models 
involving large rigid triaxiality are very close to the predictions 
of $\gamma$-soft models involving $\gamma$-fluctuations such that 
$\gamma_{rigid}$ of the former equals $\gamma_{rms}$ of the latter. 
In addition, the equivalence between $\gamma$-instability and rigid 
triaxiality with $\gamma=30^{\rm o}$ has been shown in relation to the 
O(6) limit of IBM using projection techniques \cite{Otsuka,Sugita}. 
In view of these, it is not surprising that the Z(5) predictions describe well
nuclei like $^{194}$Pt, which are known to be good examples of the O(6) 
symmetry \cite{Casten}. 

2) Taking an infinite well potential in $\beta$ (while $\gamma$ is fixed at 
$30^{\rm o}$) corresponds to a transition from a triaxial vibrator to a 
triaxial rotator \cite{MtVNPA,DavFil,DavRos},
in the same way that an infinite well potential in $\beta$ in the X(5) 
model (in which $\gamma=0^{\rm o}$ is assumed \cite{IacX5}) corresponds to 
a transition 
from a vibrator [U(5)] to a prolate rotator [SU(3)]. This point will be 
further discussed in Section 7. 

3) In view of the above, it is not surprising that the Z(5) model gives 
results compatible with earlier work on the prolate to oblate shape 
phase transition \cite{Jolie68,Jolie87}. In the earlier work 
\cite{Jolie68,Jolie87}, the $\gamma$-soft pool in which the critical point 
of the prolate to oblate transition is expected to lie, is crossed by moving 
from $\gamma=0^{\rm o}$ to $\gamma=60^{\rm o}$ in the $\beta\neq 0$ region,
i.e. away from the vibrational ($\beta=0$) regime. When moving from SU(3) 
(prolate) to $\overline{\rm SU(3)}$ (oblate) on the appropriate side of the 
extended \cite{Jolie87} Casten triangle \cite{Casten}, one then identifies 
O(6) as the critical point. In the Z(5) model the same pool is crossed 
in a different way, by fixing $\gamma=30^{\rm o}$ and moving from the 
triaxial vibrator (close to $\beta=0$) to the triaxial rotator 
(far from $\beta=0$) \cite{MtVNPA,DavFil,DavRos}.  

In Sections 2 and 3 of the present work the $\beta$-part and the $\gamma$-part
of the spectrum will be considered respectively, while B(E2) transition 
rates will be studied in Section 4. Numerical results will be reported 
in Section 5 and compared to experimental data in Section 6, while Section 
7 contains a summary of the present results and plans for further work.  

{\bf 2. The $\beta$-part of the spectrum} 

The original Bohr Hamiltonian \cite{Bohr} is
\begin{equation}\label{eq:e1}
H = -{\hbar^2 \over 2B} \left[ {1\over \beta^4} {\partial \over \partial 
\beta} \beta^4 {\partial \over \partial \beta} + {1\over \beta^2 \sin 
3\gamma} {\partial \over \partial \gamma} \sin 3 \gamma {\partial \over 
\partial \gamma} - {1\over 4 \beta^2} \sum_{k=1,2,3} {Q_k^2 \over \sin^2 
\left(\gamma - {2\over 3} \pi k\right) } \right] +V(\beta,\gamma),
\end{equation}
where $\beta$ and $\gamma$ are the usual collective coordinates, while
$Q_k$ ($k=1$, 2, 3) are the components of angular momentum and $B$ is the 
mass parameter.  

In the case in which the potential 
has a minimum around $\gamma =\pi/6$ one can write  the last term of Eq. 
(\ref{eq:e1}) in the form 
\begin{equation}\label{eq:e2} 
\sum _{k=1,2,3} {Q_k^2 \over \sin^2 \left( \gamma -{2\pi \over 3} k\right)}
\approx Q_1^2+4 (Q_2^2+Q_3^2) = 4(Q_1^2+Q_2^2+Q_3^2)-3Q_1^2. 
\end{equation}
Using this result in the Schr\"odinger equation corresponding to 
the Hamiltonian of Eq. (\ref{eq:e1}), introducing \cite{IacX5} reduced energies
 $\epsilon = 2B E /\hbar^2$ and reduced potentials $u = 2B V /\hbar^2$,  
and assuming \cite{IacX5} that the reduced potential can be separated into two 
terms, one depending on $\beta$ and the other depending on $\gamma$, i.e. 
$u(\beta, \gamma) = u(\beta) + u(\gamma)$, the Schr\"odinger equation can 
be separated into two equations 
\begin{equation} \label{eq:e3}
\left[ -{1\over \beta^4} {\partial \over \partial \beta} \beta^4 
{\partial \over \partial \beta} + {1\over 4 \beta^2} (4L(L+1)-3\alpha^2)  
+u(\beta) \right] \xi_{L,\alpha}(\beta) =\epsilon_\beta  \xi_{L,\alpha}(\beta),
\end{equation}
\begin{equation}\label{eq:e4} 
\left[ -{1\over \langle \beta^2\rangle \sin 3\gamma} {\partial \over 
\partial \gamma}\sin 3\gamma {\partial \over \partial \gamma} 
+u(\gamma)\right] \eta(\gamma) = 
\epsilon_\gamma \eta(\gamma),
\end{equation}
where $L$ is the angular momentum quantum number, $\alpha$ is the projection 
of the angular momentum on the body-fixed $\hat x'$-axis 
($\alpha$ has to be an even integer \cite{MtVNPA}),
$\langle \beta^2 \rangle$ is the average of $\beta^2$ over $\xi(\beta)$, 
and $\epsilon= \epsilon_\beta +\epsilon_\gamma$. 

The total wave function should have the form 
\begin{equation}\label{eq:e5}
\Psi(\beta,\gamma,\theta_i) = \xi_{L,\alpha}(\beta) \eta(\gamma) 
{\cal D}^L _{M,\alpha}(\theta_i), 
\end{equation}
where $\theta_i$ ($i=1$, 2, 3) are the Euler angles, ${\cal D}(\theta_i)$ 
denote Wigner functions of them, $L$ are the eigenvalues of angular 
momentum, while $M$ and $\alpha$ are the eigenvalues of the projections 
of angular momentum on the laboratory fixed $\hat z$-axis and the body-fixed 
$\hat x'$-axis respectively. 

Instead of the projection $\alpha$ of the angular momentum on the 
$\hat x'$-axis, it is customary to introduce the wobbling quantum number 
\cite{MtVNPA,BM} $n_w=L-\alpha$. Inserting $\alpha=L-n_w$ in 
Eq. (\ref{eq:e3}) one obtains 
\begin{equation} \label{eq:e6}
\left[ -{1\over \beta^4} {\partial \over \partial \beta} \beta^4 
{\partial \over \partial \beta} + {1\over 4 \beta^2} (L(L+4)+3n_w(2L-n_w))  
+u(\beta) \right] \xi_{L,n_w}(\beta) =\epsilon_\beta  \xi_{L,n_w}(\beta), 
\end{equation}
where the wobbling quantum number $n_w$ labels a series of bands 
with  $L=n_w,n_w+2,n_w+4, \dots$ (with $n_w > 0$) next to the ground state 
band (with $n_w=0$) \cite{MtVNPA}.  

In the case in which $u(\beta)$ is an infinite well potential
\begin{equation}\label{eq:e7} 
 u(\beta) = \left\{ \begin{array}{ll} 0 & \mbox{if $\beta \leq \beta_W$} \\
\infty  & \mbox{for $\beta > \beta_W$} \end{array} \right. ,  
\end{equation} 
one can use the transformation \cite{IacX5} 
$\tilde \xi(\beta) = \beta^{3/2} \xi(\beta)$, as well as the definitions 
\cite{IacX5} $\epsilon_\beta= k_\beta^2$, $z=\beta k_\beta$, in order 
to bring Eq. (\ref{eq:e6}) into the form of a Bessel equation 
\begin{equation}\label{eq:e8}
{d^2 \tilde \xi \over d z^2} + {1\over z} {d \tilde \xi \over d z} 
+ \left[ 1 - {\nu^2 \over z^2}\right] \tilde \xi=0,
\end{equation}
with 
\begin{equation}\label{eq:e9} 
\nu = {\sqrt{4L(L+1)-3\alpha^2+9}\over 2}= 
{\sqrt{L(L+4)+3n_w(2L-n_w)+9}\over 2}.   
\end{equation}
Then the boundary condition $\tilde \xi(\beta_W) =0$ 
determines the spectrum 
\begin{equation}\label{eq:e10}
\epsilon_{\beta; s,\nu} = \epsilon_{\beta; s,n_w,L} 
= (k_{s,\nu})^2, \qquad k_{s,\nu} = {x_{s,\nu}
\over \beta_W}, 
\end{equation}
and the eigenfunctions 
\begin{equation}\label{eq:e11} 
\xi_{s,\nu}(\beta) = \xi_{s,n_w,L} (\beta)= \xi_{s,\alpha,L}(\beta)= 
c_{s,\nu} \beta^{-3/2} J_\nu (k_{s,\nu} \beta), 
\end{equation}
where $x_{s,\nu}$ is the $s$th zero of the Bessel function $J_\nu(z)$, 
while the normalization constants $c_{s,\nu}$ are determined from the 
normalization condition $ \int_0^\infty \beta^4 \xi^2_{s,\nu}(\beta) 
d\beta=1$. The notation for the roots has been kept the same as in Ref. 
\cite{IacX5}, while for the energies the notation $E_{s,n_w,L}$ 
will be used. The ground state band corresponds to $s=1$, $n_w=0$.
We shall refer to the model corresponding to this solution as Z(5)
(which is not meant as a group label), in analogy to the E(5) \cite{IacE5}, 
X(5) \cite{IacX5}, and Y(5) \cite{IacY5} models.  

{\bf 3. The $\gamma$-part of the spectrum} 

The $\gamma$-part of the spectrum is obtained from Eq. (\ref{eq:e4}), 
which can be simply rewritten as
\begin{equation}\label{eq:e18} 
\left[-{1\over \langle \beta^2 \rangle} \left( {\partial^2 \over \partial 
\gamma^2} + 3{\cos 3\gamma \over \sin 3\gamma} {\partial \over \partial 
\gamma}\right) +u(\gamma)\right] \eta(\gamma) = \epsilon_\gamma \eta(\gamma). 
\end{equation} 
As already mentioned, we consider a harmonic oscillator potential having a 
minimum at $\gamma =\pi/6$, i.e. 
\begin{equation}\label{eq:e19}
u(\gamma)= {1\over 2} c \left( \gamma-{\pi \over 6}\right)^2 = 
{1\over 2} c \tilde \gamma^2, \qquad \tilde \gamma = \gamma -{\pi \over 6}.
\end{equation}
In the case of $\gamma \approx \pi/6$ the $\cos 3\gamma$ term vanishes 
and the above equation can be brought into the form 
\begin{equation}\label{eq:e20}
\left[ -{\partial^2 \over \partial \tilde \gamma^2} +{1\over 2} c 
\langle \beta^2 \rangle \tilde \gamma^2 \right] \eta(\tilde \gamma) 
= \epsilon_{\tilde \gamma} \langle \beta^2 \rangle \eta(\tilde \gamma)
\end{equation}
which is a simple harmonic oscillator equation with energy eigenvalues
\begin{equation}\label{eq:e21}
\epsilon_{\tilde \gamma} = \sqrt{2 c \over \langle \beta^2 \rangle } \left(
n_{\tilde \gamma} +{1\over 2}\right), \qquad n_{\tilde \gamma}=0,1,2,\ldots  
\end{equation}
and eigenfunctions 
\begin{equation}\label{eq:e22}
\eta_{n_{\tilde \gamma}} ({\tilde \gamma}) 
= N_{n_{\tilde \gamma}} H_{n_{\tilde \gamma}}(b \tilde \gamma) 
e^{-b^2 \tilde \gamma^2 /2}, \qquad b=\left({ c \langle \beta^2 \rangle \over 
2}\right)^{1/4},   
\end{equation}
with normalization constant 
\begin{equation}\label{eq:e23} 
N_{n_{\tilde \gamma}} = \sqrt{ b\over \sqrt{\pi} 2^{n_{\tilde \gamma}} 
n_{\tilde \gamma}! }. 
\end{equation}
Similar potentials and solutions in the $\gamma$-variable have been 
considered in \cite{Bohr,Dav24}

The total energy in the case of the Z(5) model is then
\begin{equation}\label{eq:e24}
E(s,n_w,L,n_{\tilde\gamma}) = E_0 + A (x_{s,\nu})^2 + B n_{\tilde \gamma}. 
\end{equation}

It should be noticed that in Eq. (\ref{eq:e20})
there is a latent dependence on $s$, $L$, and $n_w$ ``hidden'' in the 
$\langle \beta^2 \rangle$ term. 
The approximate separation of the $\beta$ and $\gamma$ variables
is achieved by considering an adiabatic limit, as in the X(5) case 
\cite{IacX5,Bijker}.   

{\bf 4. B(E2) transition rates} 

The quadrupole operator is given by 
\begin{equation}\label{eq:e31}
T^{(E2)}_\mu = t \beta \left[ {\cal D}^{(2)}_{\mu,0}(\theta_i)\cos\left(\gamma 
-{2\pi\over 3}\right)+{1\over \sqrt{2}}
({\cal D}^{(2)}_{\mu,2}(\theta_i)+{\cal D}^{(2)}_{\mu,-2}(\theta_i) ) 
\sin\left(\gamma -{2\pi\over 3} \right) \right],
\end{equation}
where $t$ is a scale factor, 
while in the Wigner functions the quantum number $\alpha$ appears next 
to $\mu$, and the quantity $\gamma -2\pi/3$ in the trigonometric functions 
is obtained from $\gamma-2\pi k/3$ for $k=1$, since in the present case 
the projection $\alpha$ along the body-fixed $\hat x'$-axis is used. 

For $\gamma \simeq \pi/6$ this expression is simplified into
\begin{equation}\label{eq:e32} 
T^{(E2)}_\mu = -{1\over \sqrt{2}} t \beta ({\cal D}^{(2)}_{\mu,2} (\theta_i)
+{\cal D}^{(2)}_{\mu,-2}(\theta_i)).
\end{equation}

B(E2) transition rates are given by 
\begin{equation}\label{eq:e33}
B(E2; L_i \alpha_i \to L_f \alpha_f) ={5\over 16\pi} { |\langle L_f \alpha_f
|| T^{(E2)} || L_i \alpha_i\rangle|^2 \over (2L_i+1)},  
\end{equation}
where the reduced matrix element is obtained through the Wigner-Eckart theorem
\begin{equation}\label{eq:e34} 
\langle L_f \alpha_f | T^{(E2)}_{\mu} | L_i \alpha_i\rangle =
{(L_i 2 L_f | \alpha_i \mu \alpha_f) \over \sqrt{2L_f+1}} 
\langle L_f \alpha_f || T^{(E2)} || L_i \alpha_i\rangle . 
\end{equation}

The symmetrized wave function reads 
\begin{equation}\label{eq:35}
\Psi(\beta,\gamma,\theta_i) = \xi_{s,\alpha,L}(\beta) \eta_{n_{\tilde \gamma}}
(\tilde \gamma) 
\sqrt{ 2L+1\over 16\pi^2 (1+\delta_{\alpha,0})} ({\cal D}^{(L)}_{\mu,\alpha}
+(-1)^L {\cal D}^{(L)}_{\mu,-\alpha}) ,  
\end{equation}
where the normalization factor occurs from the standard integrals 
involving two Wigner functions \cite{Edmonds} and is the same as in 
\cite{MtVNPA}. $\alpha$ has to be an even integer \cite{MtVNPA}, 
while for $\alpha=0$ it is clear that only even values of $L$ are 
allowed, since the symmetrized wave function is vanishing otherwise. 

In the calculation of the matrix elements of Eq. (\ref{eq:e34}) the integral 
over $\tilde \gamma$ leads to unity  [because of the normalization 
of $\eta(\tilde \gamma)$], the integral over $\beta$ takes the form 
\begin{equation}\label{eq:e36} 
I_\beta(s_i,L_i,\alpha_i,s_f,L_f,\alpha_f)= \int \beta 
\xi_{s_i,\alpha_i,L_i}(\beta) \xi_{s_f,\alpha_f,L_f}(\beta) \beta^4 d\beta,
\end{equation}
where the $\beta$ factor comes from Eq. (\ref{eq:e32}), and the $\beta^4$ 
factor comes from the volume element \cite{Bohr}, 
while the integral over the angles is calculated using the standard integrals 
involving three Wigner functions \cite{Edmonds}. The separation of the 
integrals occurs because $\eta(\tilde \gamma)$ does not depend 
on $\alpha$, while in $\xi(\beta)$ only even values of $\alpha$ appear.  
The final result reads 
$$ B(E2; L_i \alpha_i \to L_f \alpha_f)  = {5\over 16\pi} {t^2 \over 2}  
{1\over (1+\delta_{\alpha_i,0}) (1+\delta_{\alpha_f,0})}   $$
\begin{equation}\label{eq:e37}
\left[ (L_i 2 L_f | \alpha_i 2 \alpha_f)+ (L_i 2 L_f | \alpha_i -2 \alpha_f)
+ (-1)^{L_i} (L_i 2 L_f | -\alpha_i 2 \alpha_f) \right]^2  
I_{\beta}^2(s_i,L_i, \alpha_i,s_f,L_f,\alpha_f) .  
\end{equation}
One can easily see that the Clebsch--Gordan coefficients (CGCs) appearing 
in this equation impose a 
$\Delta \alpha=\pm 2$ selection rule. Indeed, the first CGC is nonvanishing 
only if $\alpha_i+2 = \alpha_f$, while the second CGC is nonvanishing 
only if $\alpha_i-2=\alpha_f$. The third CGC is nonvanishing only if 
$\alpha_i+\alpha_f=2$, which can be valid only in a few special cases. 
The angular part of this equation is equivalent to the results obtained 
in \cite{MtVNPA}. 

The ground state band (gsb) is characterized by $n_w=L-\alpha=0$. Therefore 
transitions within the gsb are characterized by 
$\alpha_i=L_i$ and $\alpha_f=L_f$. Normalizing the B(E2) rates to the 
lowest transition within the gsb we obtain 
\begin{equation}\label{eq:e38}
R_{g,g}(L+2\to L) = {B(E2; (L+2)_g \to L_g) \over B(E2; 2_g \to 0_g)} =
{5\over 2} {2L+1\over 2L+5} (1+\delta_{L,0}) 
{ I_{\beta}^2(1,L+2,L+2,1,L,L) \over I_{\beta}^2(1,2,2,1,0,0)},
\end{equation}
where the $(1+\delta_{L,0})$ factor comes from the fact that 
to all transitions within the gsb only the second CGC in Eq. (\ref{eq:e37})
contributes, except for the lowest one, to which both the second and 
the third terms contribute. 

The even levels of the $\gamma_1$-band are characterized by $n_w =L-\alpha=2$,
which means $\alpha=L-2$. Using the same normalization as above one 
obtains for the transitions from the even levels of the $\gamma_1$ band 
to the gsb
\begin{equation}\label{eq:e39}
R_{\gamma_{even},g}(L\to L) = {B(E2; L_\gamma \to L_g) \over B(E2; 2_g \to 0_g)}
 = {15\over (L+1)(2L+3)} (1+ \delta_{L,2})
{I_\beta^2(1,L,L-2,1,L,L) \over I_\beta^2(1,2,2,1,0,0)}, 
\end{equation}
where the $(1+\delta_{L,2})$ factor is due to the fact that for all 
transitions only the first CGC of Eq. (\ref{eq:e37}) contributes, 
except in the case of $2_\gamma\to 2_g$, in which both the first and 
the third terms contribute. 
The angular parts of Eqs. (\ref{eq:e38}) and (\ref{eq:e39}) coincide
with the results obtained in \cite{MtVPLB}.  

In a similar manner the following ratios are also derived
\begin{equation}\label{eq:e40}
R_{\gamma_{odd},g}(L\to L+1) = {B(E2; L_\gamma \to (L+1)_g) \over 
B(E2; 2_g \to 0_g)} =
{5\over L+2} {I^2(1,L,L-1,1,L+1,L+1) \over I^2(1,2,2,1,0,0)} 
\end{equation}
$$ R_{\gamma_{even} \to \gamma_{even} }(L+2\to L) = 
{B(E2; (L+2)_{\gamma} \to L_\gamma) \over B(E2; 2_g \to 0_g)} $$
\begin{equation}\label{eq:e41}
= {5 (2L-1) L(2L+1) \over 2 (2L+3) (L+2)  (2L+5) } (1+\delta_{L,2}) 
{I^2(1,L+2,L,1,L,L-2) \over I^2(1,2,2,1,0,0)}, 
\end{equation}
$$ R_{\gamma_{odd} \to \gamma_{odd}}(L+2\to L)= 
{B(E2; (L+2)_{\gamma} \to L_\gamma) \over B(E2; 2_g \to 0_g)} $$
\begin{equation}\label{eq:e42} 
={5 L (2L+1) \over 2(L+2) (2L+5)} 
{I^2(1,L+2,L+1,1,L,L-1)\over I^2(1,2,2,1,0,0)},  
\end{equation} 
$$ R_{\gamma_{odd} \to \gamma_{even} }(L\to L-1)  = 
{B(E2; L_{\gamma} \to (L-1)_\gamma) \over B(E2; 2_g \to 0_g)} $$
\begin{equation}\label{eq:e43}
= {5 (2L-3) (2L-1) \over L (L+1) (2L+1) } (1+\delta_{L,3}) 
{I^2(1,L,L-1,1,L-1,L-3) \over I^2(1,2,2,1,0,0)}.  
\end{equation}

It should be noticed that quadrupole moments vanish, because of the 
$\Delta \alpha =\pm 2$ selection rule, since in the relevant matrix 
elements of the quadrupole operator one should have $\alpha_i=\alpha_f$. 

{\bf 5. Numerical results} 

The lowest bands of the Z(5) model are given in Table 1. The notation 
$L_{s,n_w}$ is used. All levels 
are measured from the ground state, $0_{1,0}$, and are normalized to 
the first excited state, $2_{1,0}$. The ground state band is characterized 
by $s=1$, $n_w=0$, while the even and the odd levels of the $\gamma_1$-band 
are characterized by $s=1$, $n_w=2$, and $s=1$, $n_w=1$ respectively. 
The $\beta_1$-band is characterized by $s=2$, $n_w=0$. 
All these bands are characterized by $n_{\bar \gamma}=0$, and, as seen 
from Eq. (\ref{eq:e24}), are parameter free. The fact that the $\gamma_1$-band 
is characterized by $n_{\bar \gamma}=0$ is not surprising, since this is
in general the case in the framework of the rotation-vibration model 
\cite{Greiner}.     

B(E2) transition rates, normalized to the one between the two lowest states, 
B(E2;$2_{1,0}\to 0_{1,0}$), are given in Table 2. 

{\bf 6. Comparison to experiment}

Several energy levels and B(E2) transition rates predicted by the Z(5) model 
are compared in Table 3 to the corresponding experimental quantities 
of $^{194}$Pt \cite{Pt194}, which has been suggested \cite{Jolie68} to lie 
very close to the prolate to oblate critical point. Its neighbours, 
$^{192}$Pt \cite{Pt192} and $^{196}$Pt \cite{Pt196}, which demonstrate quite 
similar behaviour, 
are also shown. Not only the levels of the ground state band are well 
reproduced (below the backbending), but in addition the bandheads  of the 
$\gamma_1$-band and the $\beta_1$-band are very well reproduced, without 
involving any free parameter.  
The staggering of the theoretical levels within the $\gamma_1$-band
is quite stronger than the one seen experimentally, as it is expected 
\cite{ZC260} for models related to the triaxial rotator 
\cite{MtVNPA,DavFil,DavRos}. 

The main features of the B(E2) transition rates are also well reproduced. 
As far as the transitions from the $\gamma_1$-band to the ground state 
band are concerned, the transitions $L_{1,2}\to L_{1,0}$ are strong, 
while the transitions $(L+2)_{1,2}\to L_{1,0}$, which are forbidden 
in the Z(5) framework, are weaker by two or three orders of magnitude. 
Even the augmentation of B(E2;$2_{1,2}\to 2_{1,0}$) relative to 
B(E2;$4_{1,2}\to 4_{1,0}$), which is due to a mathematical detail, 
as explained below Eq. (\ref{eq:e39}), is very well seen experimentally. 
 
{\bf 7. Discussion}

In summary, a critical point symmetry for the prolate to oblate shape 
phase transition has been introduced by approximately separating variables 
in the Bohr Hamiltonian for $\gamma=30^{\rm o}$. The parameter free 
(up to overall scale factors) predictions of the model, called Z(5), 
are in good agreement with experimental data for $^{194}$Pt, which is 
supposed to lie close to the prolate to oblate critical point \cite{Jolie68}, 
as well as for its neighbours ($^{192}$Pt, $^{196}$Pt). 

In addition to the points made in the introduction, the following 
comments apply: 

1) The $\beta$-equation [Eq. (\ref{eq:e6})] obtained after approximately 
separating variables in the Bohr Hamiltonian is also exactly soluble 
\cite{Elliott,Rowe}
when plugging in it the Davidson potentials \cite{Dav} 
\begin{equation}\label{eq:e45}
u(\beta)= \beta^2 +{\beta_0^4\over \beta^2}, 
\end{equation}  
where $\beta_0$ is the minimum of the potential. 
In analogy to earlier work in the E(5) and X(5) frameworks \cite{varPLB},
it is expected that $\beta_0=0$ should correspond to a triaxial vibrator, 
while $\beta_0\to \infty$ should lead to a triaxial rotator
\cite{MtVNPA,DavFil,DavRos}. 

2) Using the variational procedure developed recently in the E(5) and X(5) 
frameworks \cite{varPLB}, one should be able to prove that the Z(5) 
model can be obtained from the Davidson potentials by maximizing the rate 
of change of various measures of collectivity with respect to the 
parameter $\beta_0$, thus proving that Z(5) is also the critical point 
symmetry of the transition from a triaxial vibrator to a triaxial rotator. 

Work in these directions is in progress. 

{\bf Acknowledgements} 

The authors are thankful to Jean Libert (Orsay), 
Werner Scheid (Giessen), and Victor Zamfir (Yale)  
for illuminating discussions. Support through the NATO 
Collaborative Linkage Grant PST.CLG 978799 is gratefully acknowledged.

\newpage 

\begin{table}

\caption{Energy levels of the Z(5) model (with $n_{\bar \gamma}=0$), measured 
from the $L_{s,n_w}= 0_{1,0}$ ground state and normalized to the 
$2_{1,0}$ lowest excited state. See Section 5 for further details.   
}

\bigskip

\begin{tabular}{ r r r r | r r}
\hline
$s,n_w$ & 1,0 & 1,2 & 2,0  &   & 1,1  \\
$L$     &     &     &      &$L$&      \\
\hline
  & & & & &      \\
 0 & 0.000 &       & 3.913 &   &      \\
 2 & 1.000 & 1.837 & 5.697 & 3 & 2.597\\
 4 & 2.350 & 4.420 & 7.962 & 5 & 4.634\\
 6 & 3.984 & 7.063 &10.567 & 7 & 6.869\\
 8 & 5.877 & 9.864 &13.469 & 9 & 9.318\\
10 & 8.019 &12.852 &16.646 &11 &11.989\\
12 &10.403 &16.043 &20.088 &13 &14.882\\
14 &13.024 &19.443 &23.788 &15 &18.000\\
16 &15.878 &23.056 &27.740 &17 &21.341\\
18 &18.964 &26.884 &31.942 &19 &24.905\\
20 &22.279 &30.928 &36.390 &21 &28.691\\
   &       &       &       &   &      \\
\hline
\end{tabular}
\end{table}

\newpage 

\begin{table}

\caption{B(E2) transition rates of the Z(5) model, normalized 
to the transition between the two lowest states, B(E2;$2_{1,0}\to 0_{1,0}$). 
See Sections 4 and 5 for further details. 
}

\bigskip

\begin{tabular}{ r r r | r r r | r r r}
\hline
 $L^{(i)}_{s,n_w}$ & $L^{(f)}_{s,n_w}$ & Z(5) & 
 $L^{(i)}_{s,n_w}$ & $L^{(f)}_{s,n_w}$ & Z(5) &
 $L^{(i)}_{s,n_w}$ & $L^{(f)}_{s,n_w}$ & Z(5) \\
\hline
           &           &       &           &           &       & & & \\
 $2_{1,0}$ & $0_{1,0}$ & 1.000 &           &           &       & & & \\
 $4_{1,0}$ & $2_{1,0}$ & 1.590 & $4_{1,2}$ & $2_{1,2}$ & 0.736 & 
             $5_{1,1}$ & $3_{1,1}$ & 1.235 \\   
 $6_{1,0}$ & $4_{1,0}$ & 2.203 & $6_{1,2}$ & $4_{1,2}$ & 1.031 & 
             $7_{1,1}$ & $5_{1,1}$ & 1.851 \\
 $8_{1,0}$ & $6_{1,0}$ & 2.635 & $8_{1,2}$ & $6_{1,2}$ & 1.590 & 
             $9_{1,1}$ & $7_{1,1}$ & 2.308 \\
$10_{1,0}$ & $8_{1,0}$ & 2.967 &$10_{1,2}$ & $8_{1,2}$ & 2.035 &  
            $11_{1,1}$ & $9_{1,1}$ & 2.665 \\
$12_{1,0}$ &$10_{1,0}$ & 3.234 &$12_{1,2}$ &$10_{1,2}$ & 2.394 &  
            $13_{1,1}$ &$11_{1,1}$ & 2.952 \\ 
$14_{1,0}$ &$12_{1,0}$ & 3.455 &$14_{1,2}$ &$12_{1,2}$ & 2.690 &  
            $15_{1,1}$ &$13_{1,1}$ & 3.190 \\
$16_{1,0}$ &$14_{1,0}$ & 3.642 &$16_{1,2}$ &$14_{1,2}$ & 2.938 &  
            $17_{1,1}$ &$15_{1,1}$ & 3.392 \\
$18_{1,0}$ &$16_{1,0}$ & 3.803 &$18_{1,2}$ &$16_{1,2}$ & 3.151 &  
            $19_{1,1}$ &$17_{1,1}$ & 3.566 \\
$20_{1,0}$ &$18_{1,0}$ & 3.944 &$20_{1,2}$ &$18_{1,2}$ & 3.335 &  & & \\
           &           &       &           &           &       &  & & \\
\hline
  &  & & & & & & &      \\
 $2_{1,2}$ & $2_{1,0}$ & 1.620 & $3_{1,1}$ & $4_{1,0}$ & 1.243 & 
           $3_{1,1}$ & $2_{1,2}$ & 2.171 \\    
 $4_{1,2}$ & $4_{1,0}$ & 0.348 & $5_{1,1}$ & $6_{1,0}$ & 0.972 & 
           $5_{1,1}$ & $4_{1,2}$ & 1.313 \\
 $6_{1,2}$ & $6_{1,0}$ & 0.198 & $7_{1,1}$ & $8_{1,0}$ & 0.808 & 
           $7_{1,1}$ & $6_{1,2}$ & 1.260 \\ 
 $8_{1,2}$ & $8_{1,0}$ & 0.129 & $9_{1,1}$ &$10_{1,0}$ & 0.696 & 
           $9_{1,1}$ & $8_{1,2}$ & 1.164 \\
$10_{1,2}$ &$10_{1,0}$ & 0.092 &$11_{1,1}$ &$12_{1,0}$ & 0.614 &
           $11_{1,1}$ &$10_{1,2}$ & 1.069 \\
$12_{1,2}$ &$12_{1,0}$ & 0.069 &$13_{1,1}$ &$14_{1,0}$ & 0.551 &  
           $13_{1,1}$ &$12_{1,2}$ & 0.984 \\
$14_{1,2}$ &$14_{1,0}$ & 0.054 &$15_{1,1}$ &$16_{1,0}$ & 0.507 &  
           $15_{1,1}$ &$14_{1,2}$ & 0.910 \\
$16_{1,2}$ &$16_{1,0}$ & 0.043 &$17_{1,1}$ &$18_{1,0}$ & 0.459 &  
           $17_{1,1}$ &$16_{1,2}$ & 0.846 \\
$18_{1,2}$ &$18_{1,0}$ & 0.035 &$19_{1,1}$ &$20_{1,0}$ & 0.425 &  
           $19_{1,1}$ &$18_{1,2}$ & 0.790 \\
$20_{1,2}$ &$20_{1,0}$ & 0.030 & & & & & & \\
  &   &       &       &       & & & & \\  
\hline
\end{tabular}
\end{table}

\newpage 

\begin{table}

\caption{Comparison of the Z(5) predictions for energy levels (left part) 
and B(E2) transition rates (right part) to experimental data for 
$^{192}$Pt \cite{Pt192}, $^{194}$Pt \cite{Pt194}, and $^{196}$Pt \cite{Pt196}. 
See Section 6 for further discussion.  
}

\bigskip

\begin{tabular}{ r r r r r | r r r r r r }
\hline
$L_{s,n_w}$ & Z(5) & $^{192}$Pt & $^{194}$Pt & $^{196}$Pt &
$L^{(i)}_{s,n_w}$ & $L^{(f)}_{s,n_w}$ & Z(5) & $^{192}$Pt & $^{194}$Pt & 
$^{196}$Pt \\
\hline
 &  &  &  &  &  &  &  &  &  &  \\
$4_{1,0}$  & 2.350 & 2.479 & 2.470 & 2.465 &
$4_{1,0}$ & $2_{1,0}$ & 1.590 & 1.559 & 1.724 & 1.476 \\
$6_{1,0}$  & 3.984 & 4.314 & 4.299 & 4.290 &
          &           &       &       &       &       \\
$8_{1,0}$  & 5.877 & 6.377 & 6.392 & 6.333 &
$4_{1,2}$ & $2_{1,2}$ & 0.736 &       & 0.446 & 0.715 \\
$10_{1,0}$ & 8.019 & 8.624 &       & 8.558 &
$6_{1,2}$ & $4_{1,2}$ & 1.031 &       &       & 1.208 \\
           &       &       &       &       &
          &           &       &       &       &       \\
$2_{1,2}$  & 1.837 & 1.935 & 1.894 & 1.936 &
$3_{1,1}$ & $2_{1,2}$ & 2.171 & 1.786 &       &       \\
$4_{1,2}$  & 4.420 & 3.795 & 3.743 & 3.636 &
          &           &       &       &       &       \\
$6_{1,2}$  & 7.063 & 5.905 & 5.863 & 5.644 &
$2_{1,2}$ & $0_{1,0}$ & 0.000 & 0.009 & 0.006 & 0.0004\\
           &       &       &       &       & 
$2_{1,2}$ & $2_{1,0}$ & 1.620 & 1.909 & 1.805 &       \\
$3_{1,1}$  & 2.597 & 2.910 & 2.809 & 2.854 &
$4_{1,2}$ & $2_{1,0}$ & 0.000 &       & 0.004 & 0.014 \\
$5_{1,1}$  & 4.634 & 4.682 & 4.563 & 4.526 &
$4_{1,2}$ & $4_{1,0}$ & 0.348 &       & 0.406 &       \\   
$7_{1,1}$  & 6.869 & 6.677 &       &       & 
$6_{1,2}$ & $4_{1,0}$ & 0.000 &       &       & 0.012 \\
           &       &       &       &       &   &  &  &  &  &  \\
$0_{2,0}$  & 3.913 & 3.776 & 3.858 & 3.944 &   &  &  &  &  &  \\
 &  &  &  &  &   &  &  &  &  &  \\
\hline
\end{tabular}
\end{table}

\end{document}